\documentclass[preprint,showpacs,showkeys,pra]{revtex4}
\usepackage{epsf}
\usepackage{amsmath,amsthm,amssymb}
\usepackage{dcolumn}
\usepackage{bm}
\everymath{\displaystyle}

\begin{document}           

\title{Potential for measurement of the tensor polarizabilities of
nuclei in storage rings by the frozen spin method}

\author{Alexander J. Silenko} 
\affiliation{Institute of Nuclear Problems, Belarusian State
University, Minsk 220030, Belarus}

\date{\today}

\begin {abstract}
The frozen spin method can be effectively used for a
high-precision measurement of the tensor electric and magnetic
polarizabilities of the deuteron and other nuclei in storage
rings. For the deuteron, this method would provide the
determination of the deuteron's polarizabilities with absolute
precision of order of $10^{-43}$ cm$^3$.

\end{abstract}
\maketitle

\section{INTRODUCTION}

Tensor electric and magnetic polarizabilities are important
properties of the deuteron and other nuclei defined by spin
interactions of nucleons. Their measurement provides a good
possibility to examine the theory of spin-dependent nuclear
forces. Methods for determining these important electromagnetic
properties of the deuteron based on the appearance of interactions
quadratic in the spin have been proposed by V. Baryshevsky and
co-workers \cite{Bar1,Bar4,Bar3}. Additional investigations have
been performed in Refs. \cite{PRC,PRC2008}.

Interactions quadratic in the spin and proportional to the tensor
electric and magnetic polarizabilities affect spin dynamics. When
an electric field in the particle rest frame oscillates at the
resonant frequency, an effect similar to the nuclear magnetic
resonance takes place. This effect stimulates the buildup of the
vertical polarization (BVP) of the deuteron beam
\cite{Bar1,Bar4,Bar3}. General formulas describing the BVP caused
by the tensor electric polarizability of the deuteron in storage
rings (the Baryshevsky effect) have been derived in Ref.
\cite{PRC}. The problem of influence of the tensor electric
polarizability on spin dynamics in such a deuteron
electric-dipole-moment experiment in storage rings has been
investigated \cite{PRC}. It has been proved that doubling the
resonant frequency used in this experiment dramatically amplifies
the Baryshevsky effect and provides the opportunity to make
high-precision measurements of the deuteron's tensor electric
polarizability \cite{PRC}.

The tensor magnetic polarizability, $\beta_T$, produces the spin
rotation with two frequencies instead of one, beating with a
frequency proportional to $\beta_T$, and causes transitions
between vector and tensor polarizations \cite{Bar4,Bar3}. In Ref.
\cite{PRC2008}, the existence of these effects has been confirmed
and a detailed calculation of deuteron spin dynamics in storage
rings has been carried out. The use of the matrix Hamiltonian
derived in Ref. \cite{PRC} is very helpful for calculating general
formulas describing the evolution of the spin. Significant
improvement in the precision of possible experiments can be
achieved if initial deuteron beams are tensor-polarized
\cite{PRC,PRC2008}.

The frozen spin method \cite{NSPC,EDM} provides another
possibility to measure the tensor polarizabilities of the deuteron
and other nuclei. This method remains the spin orientation
relatively the momentum direction almost unchanged. In the present
work, we also analyze additional advantages ensured by the use of
tensor-polarized beams and compute the related spin evolution.

The system of units $\hbar=c=1$ is used.

\section{General equations}

The traditional quantum mechanical approach
(see Ref. \cite{RF}) uses the matrix
Hamiltonian equation and the matrix Hamiltonian $H$ for
determining an evolution of the spin wave function:
\begin{equation}
 \begin{array}{c}
 i \frac{d\Psi}{dt}=H\Psi, ~~~ \Psi=\left(\begin{array}{c}
C_{1}(t)
 \\ C_{0}(t) \\ C_{-1}(t) \end{array}\right).
 \end{array}\label{eq19t}\end{equation}
The matrix Hamiltonian $H$ coincides with the Hamilton operator
${\cal H}$ expressed in matrix form. This coincidence results from
the fact that the Hamilton operator considered in Refs.
\cite{PRC,PRC2008} is independent of coordinates.

Three-component wave function $\Psi$ which is similar to a spinor
consists of the amplitudes $C_{i}(t)$ characterizing states with
definite spin projections onto the preferential direction ($z$
axis). Correction to the Hamilton operator caused by the tensor
polarizabilities has the form \cite{PRC}
\begin{equation} \begin{array}{c}
V=-\frac{\alpha_T}{\gamma}(\bm S\cdot\bm
E')^2-\frac{\beta_T}{\gamma}(\bm S\cdot\bm B')^2,
\end{array} \label{eqTre} \end{equation}
where $\alpha_T$ is the tensor electric polarizability, $\gamma$
is the Lorentz factor, and $\bm E'$ and $\bm B'$ are the electric
and magnetic fields in the rest frame of the deuteron.

The use of the rotating frame is very helpful. The particle in
this frame is localized and ideally is at rest. Therefore, we can
direct the $x$ and $y$ axes in this frame along the radial and
longitudinal axes, respectively. This procedure is commonly used
(see Ref. \cite{PRC} and references therein) and results in the
simplest forms of spin matrices:
 \begin{equation}
\begin{array}{c}
S_\rho=\frac{1}{\sqrt{2}}\left(\begin{array}{ccc} 0 &
 1 & 0\\ 1 & 0 & 1\\ 0 & 1 & 0
 \end{array}\right), ~~~ S_\phi=\frac{i}{\sqrt{2}}\left(\begin{array}{ccc} 0 &
 -1 & 0\\ 1 & 0 & -1\\ 0 & 1 & 0
 \end{array}\right),\\ S_z=\left(\begin{array}{ccc} 1 &
 0 & 0\\ 0 & 0 & 0\\ 0 & 0 & -1
 \end{array}\right).
 \end{array}\label{eq20}\end{equation}

The Hamiltonian operator is defined by \cite{PRC}
\begin{equation} \begin{array}{c}
{\cal H}={\cal H}_0+\bm S\cdot\bm\omega_a+V,
\end{array} \label{eqapp} \end{equation}
where $\bm\omega_a$ is the angular velocity of the spin precession
relatively to the momentum direction (g$-$2 precession).

In the considered case, the expression of $\bm E'$ and $\bm B'$ in
terms of the lab fields which are unprimed has the form
\begin{equation} \begin{array}{c}
\bm E'=\gamma(E_\rho+\beta_\phi B_z)\bm e_\rho, ~~~ \bm
B'=\gamma(\beta_\phi E_\rho+B_z)\bm e_z,
\end{array} \label{eqTr} \end{equation}
where $\beta_\phi=\bm\beta\cdot \bm e_\phi\equiv \bm v\cdot \bm
e_\phi/c$.

When the frozen spin method is used, the quantity $\bm\omega_a$ is
very small and the fields satisfy the following relation
\cite{EDM,N84}:
\begin{equation} \begin{array}{c}
E_\rho=\frac{a\beta_\phi\gamma^2}{1-a\beta^2\gamma^2} B_z.
\end{array} \label{eqc} \end{equation}
For the deuteron, $a\equiv(g-2)/2=-0.143$. Therefore,
\begin{equation} \begin{array}{c}
V=-\frac{\gamma B_z^2}{(1-a\beta^2\gamma^2)^2}\left[\alpha_T(1+a)^2\beta^2
S_\rho^2+\beta_TS_z^2\right].
\end{array} \label{eqTrI} \end{equation}

The matrix Hamiltonian has the form \cite{PRC}
\begin{equation}
 \begin{array}{c}
 H=\left(\begin{array}{ccc} E_{0}\!+\!\omega_0\!+\!{\cal A}\!+\!{\cal B} & 0 & {\cal A} \\
0 & E_{0}\!+\!2{\cal A} & 0 \\
{\cal A} & 0 & E_{0}\!-\!\omega_0\!+\!{\cal A}\!+\!{\cal B}
 \end{array}\right),
 \end{array}\label{eqMH}\end{equation}
where $E_0$ is the zero energy level,
$\omega_0=\left(\omega_a\right)_z$,
\begin{equation}
\begin{array}{c}
{\cal A}=-\alpha_T\frac{(1+a)^2\beta^2\gamma B_z^2}{2(1-a\beta^2\gamma^2)^2},
~~~ {\cal B}=-\beta_T\frac{\gamma B_z^2}{(1-a\beta^2\gamma^2)^2}.
\end{array}\label{eqMHt}\end{equation}

Eqs. (\ref{eqMH}),(\ref{eqMHt}) are basic equations defining
dynamics of the deuteron spin in storage rings when the frozen
spin method is used.

\section{Evolution of vector polarization of deuteron beam}

In Ref. \cite{PRC2008}, off-diagonal components of Hamiltonian
(\ref{eqMHt}) was not taken into account, because their effect on
the rotating spin did not satisfy the resonance condition. These
components cannot, however, be neglected in the considered case
because the resonant frequency $\omega_0$ can be very small.

The best conditions for a measurement of the tensor
polarizabilities of the deuteron and other nuclei can be achieved
with the use of tensor-polarized initial beams. In this case, we
may confine ourselves to the consideration of a zero projection of
the deuteron spin onto the preferential direction. When this
direction is defined by the spherical angles $\theta$ and $\psi$
\cite{foo}, the initial polarization is given by
\begin{equation}\begin{array}{c}
\bm P(0)=0, ~~~
P_{\rho\rho}(0)=1-3\sin^2{\theta}\cos^2{\psi}, \\
P_{\phi\phi}(0)=1-3\sin^2{\theta}\sin^2{\psi}, ~~~
P_{zz}(0)=1-3\cos^2{\theta}, \\ 
P_{\rho\phi}(0)=-\frac32\sin^2{\theta}\sin{(2\psi)}, \\ P_{\rho
z}(0)\!=\!-\frac32\sin{(2\theta)}\cos{\psi}, ~~~ P_{\phi
z}(0)\!=\!-\frac32\sin{(2\theta)}\sin{\psi}.
\end{array}\label{intvi}
\end{equation}

In this case, the general equation describing the evolution of the
polarization vector has the form
\begin{equation}
\begin{array}{c}
P_\rho(t)=\sin{(2\theta)}\biggl\{\Bigl[\cos{(\omega't)}\sin{\psi}
\\
+\frac{\omega_0}{\omega'}\sin{(\omega't)}\cos{\psi}\Bigr]\sin{(bt)}\!+\!
\frac{{\cal A}}{\omega'}\sin{(\omega't)}\cos{(bt)}\sin{\psi}\biggr\},\\
P_\phi(t)=\sin{(2\theta)}\biggl\{\Bigl[-\cos{(\omega't)}\cos{\psi}
\\
+\frac{\omega_0}{\omega'}\sin{(\omega't)}\sin{\psi}\Bigr]\sin{(bt)}
\!+\!\frac{{\cal A}}{\omega'}\sin{(\omega't)}\cos{(bt)}\cos{\psi}\biggr\},\\
P_{z}(t)=-\frac{2{\cal
A}}{\omega'}\sin^2{\theta}\sin{(\omega't)}\Bigl[\cos{(\omega't)}
\sin{(2\psi)} \\
+\frac{\omega_0}{\omega'}\sin{(\omega't)}\cos{(2\psi)}\Bigr],
\end{array}
\label{prop}
\end{equation}
where
\begin{equation}
\omega'=\sqrt{\omega_0^2+{\cal A}^2}, ~~~ b={\cal B}-{\cal A}.
\label{eqb}
\end{equation}
When the frozen spin method is used,
\begin{equation}
b=-\frac{\gamma
B_z^2}{(1-a\beta^2\gamma^2)^2}\left[\beta_T-\frac{1}{2}\alpha_T(1+a)^2\beta^2\right].
\label{eqbf}
\end{equation}

As a rule, we can neglect ${\cal A}^2$ as compared with
$\omega_0^2$ and use the approximation $bt\ll1$. In this case,
\begin{equation}
\begin{array}{c}
P_\rho(t)=\sin{(2\theta)}\left[bt\sin{(\omega_0t+\psi)}+ 
\frac{{\cal A}}{\omega_0}\sin{(\omega_0t)}\sin{\psi}\right],\\
P_\phi(t)=\sin{(2\theta)}\!\left[-bt\cos{(\omega_0t+\psi)}\!+\!
\frac{{\cal A}}{\omega_0}\sin{(\omega_0t)}\cos{\psi}\right],\\
P_{z}(t)=-\frac{2{\cal
A}}{\omega_0}\sin^2{\theta}\sin{(\omega_0t)}\sin{(\omega_0t+2\psi)}.
\end{array}
\label{prp}
\end{equation}

When the initial deuteron beam is vector-polarized and the
direction of its polarization is defined by the spherical angles
$\theta$ and $\psi$, 
\begin{equation}\begin{array}{c}
P_{\rho}(0)=\sin{\theta}\cos{\psi}, ~~~
P_{\phi}(0)=\sin{\theta}\sin{\psi}, \\ P_{z}(0)=\cos{\theta}, ~~~
P_{\rho\rho}(0)=\frac12\left(3\sin^2{\theta}\cos^2{\psi}-1\right),\\
P_{\phi\phi}(0)=\frac12\left(3\sin^2{\theta}\sin^2{\psi}-1\right),\\
P_{\rho\phi}(0)=\frac34\sin^2{\theta}\sin{(2\psi)}.
\end{array} \label{eq4} \end{equation}
Such a polarization (with $\theta=\pi/2$) will be used in the
planned deuteron electric-dipole-moment (EDM) experiment
\cite{dEDM}. The EDM manifests in
an appearance of a 
vertical component of the polarization vector.

The evolution of this component defined by the tensor
polarizabilities of the deuteron is given by
\begin{equation}
\begin{array}{c}
P_{z}(t)=\left[1-\frac{2{\cal
A}^2}{{\omega'}^2}\sin^2{(\omega't)}\right]\cos{\theta}
\\+\frac{{\cal
A}}{\omega'}\sin^2{\theta}\sin{(\omega't)}\Bigl[\cos{(\omega't)}
\sin{(2\psi)} \\
+\frac{\omega_0}{\omega'}\sin{(\omega't)}\cos{(2\psi)}\Bigr].
\end{array}
\label{propv}
\end{equation}
The tensor magnetic polarizability does not influence on $P_{z}$.

In the same approximation as before,
\begin{equation}
\begin{array}{c}
P_{z}(t)=\cos\theta+\frac{{\cal
A}}{\omega_0}\sin^2{\theta}\sin{(\omega_0t)}\sin{(\omega_0t+2\psi)}.
\end{array}
\label{aprop}
\end{equation}

\section{Discussion and summary}

Experimental conditions needed for the measurement of the tensor
polarizabilities and the EDMs of nuclei in storage rings
\cite{EDM,dEDM} are similar. Eq. (\ref{eqc}) shows that the radial
electric field should be sufficiently strong in order to eliminate
the effect of the vertical magnetic field on the spin. As a
result, the frozen spin method provides a weaker magnetic field
than other methods. This factor is negative because the evolution
of the spin caused by both the tensor polarizabilities and the
EDMs strongly depends on $B_z$. Nevertheless, the Storage Ring EDM
collaboration considers the frozen spin method to be capable to
detect the deuteron EDM of order of $10^{-29} e \cdot\,$cm.
Another method for searching for the deuteron EDM in storage rings
is the resonance method developed in Ref. \cite{OMS}. This method
is based on a strong vertical magnetic field and an oscillatory
resonant longitudinal electric field. The use of the resonance
method for the measurement of the tensor electric polarizability
has been proposed in Refs. \cite{Bar1,Bar4,Bar3}. This method may
provide higher sensitivity than the frozen spin method (see Ref.
\cite{PRC} for comparing the results). However, the realization of
the resonance method seems to be more difficult. Evidently, a
strong restriction of the spin rotation considerably simplifies a
detection of weak spin-dependent effects. In addition, this
restriction facilitates adjusting a storage ring lattice which is
essentially based on monitoring a behavior of the spin
\cite{EDM,dEDM,OMS}.

We can evaluate a precision of measurement of the tensor
polarizabilities of the deuteron via its comparison with the
expected sensitivity of the deuteron EDM experiment.

Evidently, the tensor electric polarizability can in principle
imitate the presence of the EDM. The exact equation of spin motion
in storage rings with allowance for EDMs of nuclei has been
derived in Ref. \cite{N84}. In the considered case, the angular
velocity of spin rotation is equal to
\begin{eqnarray}
\bm\omega_a=\omega_0\bm e_z+{\cal C}\bm e_\rho, ~~~ {\cal
C}=-\frac{e\eta}{2m}\cdot\frac{1+a}{1-a\beta^2\gamma^2}\beta_\phi
B_z, \label{eq8}\end{eqnarray} where $\eta=2dm/(eS)$ is the factor
similar to $g$ factor for the magnetic moment. $d$ is the EDM.

When the tensor polarizabilities are not taken into account, the
spin rotates about the direction
$$\bm e_z'=\frac{{\cal C}}{\omega'}\bm e_\rho+\frac{\omega_0}{\omega'}\bm e_z$$
with the angular frequency $\omega'=\sqrt{\omega_0^2+{\cal C}^2}$.

When the initial polarization of the beam is given by Eq.
(\ref{eq4}), the polarization vector is equal to
\begin{equation}
\begin{array}{c}
P_{\rho}(t)=\frac{{\omega_0\cal
C}}{{\omega'}^2}\left[1-\cos{(\omega't)}\right]\cos{\theta}\\
+\!\left[1\!-\!\frac{2\omega_0^2}{{\omega'}^2}\sin^2{\frac{\omega't}{2}}\right]\sin{\theta}
\cos{\psi}-\frac{\omega_0}{\omega'}\sin{(\omega't)}\sin{\theta}\sin{\psi},\\
P_{\phi}(t)=\sin{(\omega't)}\left(\frac{\omega_0}{\omega'}
\sin{\theta}\cos{\psi}-\frac{{\cal
C}}{\omega'}\cos{\theta}\right)\\+\cos{(\omega't)}
\sin{\theta}\sin{\psi},\\
P_{z}(t)=\left[1-\frac{2{\cal
C}^2}{{\omega'}^2}\sin^2{\frac{\omega't}{2}}\right]\cos{\theta}
\\+\frac{{\omega_0\cal
C}}{{\omega'}^2}\left[1-\cos{(\omega't)}\right]\sin{\theta}
\cos{\psi}+\frac{{\cal
C}}{\omega'}\sin{(\omega't)}\sin{\theta}\sin{\psi}.
\end{array}
\label{propedm}
\end{equation}
If we neglect terms of order of ${\cal C}^2$, the vertical
component of the polarization vector takes the form
\begin{equation}
\begin{array}{c}
P_{z}(t)=\cos{\theta}+\frac{2{\cal
C}}{\omega_0}\sin{\theta}\sin{\frac{\omega_0t}{2}}\sin{\frac{\omega_0t+2\psi}{2}}.
\end{array}
\label{appredm}
\end{equation}

While Eqs. (\ref{aprop}) and (\ref{appredm}) are similar, the
effects of the tensor electric polarizability and the EDM have
different angular dependencies and can be properly separated.

For the considered experimental conditions \cite{dEDM}, the
sensitivity to the EDM of $1\times10^{-29} e \cdot\,$cm
corresponds to measuring the tensor electric polarizability with
the accuracy $\delta\alpha_T\approx 5\times10^{-42}$ cm$^3$.

There are three independent theoretical predictions for the value
of the tensor electric polarizability of the deuteron, namely
$\alpha_T=-6.2\times10^{-41}$ cm$^3$ \cite{CGS},
$-6.8\times10^{-41}$ cm$^3$ \cite{JL}, and $3.2\times10^{-41}$
cm$^3$ \cite{FP}. The first two values are very close to each
other but they do not agree with the last result. The theoretical
estimate for the tensor magnetic polarizability of deuteron is
$\beta_T=1.95\times10^{-40}$ cm$^3$ \cite{CGS,JL}.

We can therefore conclude that the expected sensitivity of the
deuteron EDM experiment allows to measure the tensor electric
polarizability with absolute precision
$\delta\alpha_T\approx5\times10^{-42}$ cm$^3$ which corresponds to
the relative precision of order of $10^{-1}$. This estimate is
made for the vector-polarized initial beam. However, the best
sensitivity in the measurement of $\alpha_T$ can be achieved with
the use of a tensor-polarized initial beam. When the vector
polarization of such a beam is zero, any spin rotation does not
occur. In this case, there are no related systematic errors caused
by the radial magnetic field and some other reasons. In the
general case, such systematic errors are proportional to a
residual vector polarization of the beam. This advantage leads to
a sufficient increase in experimental accuracy \cite{PRC,PRC2008}.
In this case, our preliminary estimate of experimental accuracy is
$\delta\alpha_T\sim 10^{-43}$ cm$^3$.

The frozen spin method can also be successively used for the
measurement of the tensor magnetic polarizability. Eqs.
(\ref{prop})--(\ref{prp}) show that the preferential direction of
initial tensor polarization is defined by $\theta=\pi/2$ and
$\theta=\pi/4$ for measuring the tensor electric and magnetic
polarizabilities, respectively. In the latter case, the horizontal
components of the polarization vector should be measured. Due to a
restriction of spin rotation in the horizontal plane, the
achievable absolute precision of measurement of the tensor
magnetic polarizability of the deuteron is of the same order
($\delta\beta_T\sim 10^{-43}$ cm$^3$). A comparison with the
theoretical estimate \cite{CGS,JL} shows that the relative
precision of measurement of this quantity can be rather high
($\delta\beta_T/\beta_T\sim 10^{-3}$).

All above derived formulas are applicable to any spin-1 nuclei.
Moreover, the evolution of the polarization vector defined by spin
tensor effects has to be identical for nuclei with any spin
$S\geq1$ despite difference of spin matrices. This statement
follows from the fact that quantum mechanical equations describing
spin dynamics should agree with classical spin physics and
therefore should not explicitly depend on $S$.

Thus, the frozen spin method can be effectively used for the
high-precision determination of the tensor electric and magnetic
polarizabilities of the deuteron and other nuclei.

\section*{ACKNOWLEDGMENT}

The author is grateful to V. G. Baryshevsky for helpful
discussions. This work was supported by the Belarusian Republican
Foundation for Fundamental Research (grant No. $\Phi$08D-001).

\end{document}